\documentstyle[aps,epsf,12pt]{revtex}

\begin{document}

\title{\bf Non-Equilibrium Phase Transition in Rapidly Expanding Matter}

\author{I.N. Mishustin\\ 
{\it The Niels Bohr Institute, Blegdamsvej 17, DK-2100 
Copenhagen {\O}, Denmark;}\\
{\it The Kurchatov Institute, Russian Research Center, Moscow 
123182, Russia}}

\maketitle

\begin{abstract} 

Non-equilibrium features of a first order phase transition from the 
quark-gluon plasma to a hadronic gas in relativistic heavy-ion collisions 
are discussed. It is demonstrated that strong collective expansion may 
lead to the fragmentation of the plasma phase into droplets surrounded 
by undersaturated hadronic gas. Subsequent hadronization of droplets will 
generate strong non-statistical fluctuations in the hadron rapidity 
distribution in individual events. The strongest fluctuations are expected 
in the vicinity of the phase transition threshold.

\end{abstract}
\hspace{1cm}
PACS numbers: 12.38.Mh; 12.39.Fe; 25.75.-q

\vspace{0.5cm}

The main goal of present and future experiments with relativistic heavy ions 
is to produce and study in the laboratory a new form of strongly interacting 
matter, the Quark-Gluon Plasma (QGP). Due to the confinement of color charges,
only colorless hadronic final states can be observed experimentally. 
Therefore, QGP properties can be studied only indirectly through the final 
hadron distributions or by penetrating electromagnetic probes. Many QGP 
signatures have been proposed in the recent years, particularly ones which 
assume an equilibrium phase transition between QGP and hadronic gas. 

The phase structure of QCD is not yet fully understood. Reliable lattice
calculations exist only for baryon-free matter where they predict a 
second order phase transition or crossover at $T_c\approx 160$ MeV. Recent 
calculations using different models \cite{Moc,Raj,Jac,Car} reveal the 
possibility of a first order phase transition at large baryon chemical 
potentials and moderate temperatures. The predicted phase diagram in the
$(T,\mu)$ plane contains a first order transition line (below called 
the critical line) terminated at $T\approx 120$ MeV by a (tri)critical point 
\cite{Raj,Jac}. Possible signatures of this point in heavy-ion collisions 
are discussed in Ref. \cite{Ste}. Under certain non-equilibrium conditions, 
a first order transition is also predicted for baryon-free matter \cite{Sat}.

A striking feature of relativistic heavy-ion collisions, confirmed
in many experiments (see e.g. \cite{Bra}), is a very strong collective
expansion of matter. The applicability of equilibrium concepts for
describing phase transitions under such conditions becomes questionable. 
The goal of this paper is to demonstrate that non-equilibrium phase 
transitions in rapidly expanding matter can lead to interesting 
phenomena which, in a certain sense, can be even easier to observe.

To make the discussion below more concrete, I adopt a picture of 
the chiral phase transition for which the mean chiral field $\Phi=(\sigma,{\bf 
\pi})$ serves as an order parameter. It is assumed that the theory respects 
chiral symmetry, which is spontaneously broken in the vacuum where 
$\sigma=f_{\pi}$, ${\bf \pi}=0$. 
The effective thermodynamic potential $\Omega(T,\mu;\Phi)$ depends,
besides $\Phi$, on temperature $T$ and baryon chemical potential $\mu$.  
Since explicit symmetry breaking terms are supposed to be small, to a good 
approximation $\Omega$ is a function of $\Phi^2=\sigma^2+{\bf \pi}^2$.

The schematic behaviour of $\Omega(T,\mu;\Phi)$ as a function of the order
parameter field $\sigma$ at $\pi=0$ is shown in Fig. 1. 
The curves from bottom to top correspond to different stages of the 
isentropic expansion of homogeneous matter. 
Each curve represents a certain point on the ($T,\mu$) trajectory. 
The minima of $\Omega$ correspond to the stable or metastable states of 
matter under the condition of thermodynamical equilibrium, where the pressure 
is $P=-\Omega_{min}$. 
The figure is based on the calculations within the linear $\sigma$-model
with constituent quarks \cite{Moc}, which predicts a rather weak first order 
phase 
transition. A similar structure of $\Omega(T,\mu;\Phi)$ but, possibly, with 
a stronger phase transition is predicted by the NJL model \cite{Raj} and by 
the random matrix model \cite{Jac}. The discussion below is quite general.
 
Assume that at some early stage of the reaction thermal (but not 
necessarily chemical) equilibrium is established and partonic matter 
is in a ``high energy density'' phase Q. This state corresponds to the 
absolute minimum of $\Omega$ with the order parameter 
close to zero, $\sigma\approx 0$, ${\bf \pi}\approx 0$, and chiral symmetry 
restored (curve 1). Due to a very high internal pressure, Q matter will 
expand and cool down.
At some stage a metastable minimum appears in $\Omega$ 
at a finite value of $\sigma$ corresponding to a ``low energy density''
phase H, in which chiral symmetry is spontaneously broken. 
At some later time, the critical line in the ($T,\mu$) plane is crossed 
where the Q and H minima have equal depths, i.e. $P_{\rm H}=P_{\rm Q}$ 
(curve 2). At later times the H phase 
becomes more favourable (curve 3), but the two phases are still 
separated by the potential barrier. 
If the expansion of the Q phase continues until the barrier vanishes 
(curve 4), the system will find itself in an absolutely unstable state at 
a maximum of the thermodynamic potential. Therefore, it will  
freely roll down into the lower energy state corresponding to the H phase.
This situation is known as a spinodal instability.  

According to the standard theory of homogeneous nucleation \cite{Kap}, 
supercritical bubbles of the H phase appear only below the critical 
line. Under condition of thermal equilibrium between 
the two phases, the supercritical bubbles can only grow through the 
conversion of new portions of the Q matter into the H phase on the bubble 
boundary. The bubble growth is then limited by a small viscosity of the
Q phase resulting in a slow dissipation of the latent heat \cite{Kap}. 
Therefore, a certain degree of supercooling is needed in order to  convert 
a significant fraction of the Q matter into the H phase in the form of 
nucleation bubbles \cite{Kap,Zab}.

In rapidly expanding matter the nucleation picture might be very different. 
Let us consider first an isotropically expanding system with 
the collective velocity field which follows the Hubble law locally,
$v(r)=H\cdot r$.
The Hubble ``constant'' $H$ may in general be a function of time, 
e.g. $H\sim 1/t$. 
Obviously, the same velocity field is seen from any local rest frame 
comoving with the matter. Suppose that a bubble of 
the H phase has formed in the expanding Q matter because of a statistical 
fluctuation. The change in thermodynamic potential of the system can be 
decomposed into three parts, 
\begin{equation} \label{delta}
\Delta\Omega= \Delta\Omega_{bulk}+ \Delta\Omega_{surf}+ \Delta\Omega_{kin}~.  
\end{equation}
In the thin-wall approximation the bulk and surface terms are expressed 
through the bubble radius $R$ in a standard way,
\begin{equation} \label{busu}
\Delta\Omega_{bulk}=-\frac{4\pi}{3}R^3\left(P_{\rm H}-P_{\rm Q}\right)~,~~~
\Delta\Omega_{surf}=4\pi R^2\gamma~,
\end{equation}
where $P_{\rm H}$ and $P_{\rm Q}$ are the pressures 
of the bulk H and Q phases, $\gamma$ is the effective surface tension.  
The last term in Eq. (\ref{delta}) accounts for the change in the 
local kinetic energy of expanding matter. In the same approximation it can 
be evaluated as
\begin{equation} \label{kin}
\Delta\Omega_{kin}=\frac{1}{2}\int_0^R 4\pi r^2dr{\cal E}(r)v^2(r)
\approx-\frac{2\pi}{5}R^5\Delta {\cal E} H^2~,
\end{equation}
where $\Delta {\cal E}\equiv{\cal E}_{\rm Q}-{\cal E}_{\rm H}$ is the 
difference 
of energy (more exactly, enthalpy) densities of the two bulk phases. 
It is evident that this term is negative, since the dense Q phase is replaced
in the bubble by the dilute H phase (typically, 
${\cal E}_{\rm Q}\gg{\cal E}_{\rm H}$). Thus, the bubble formation is 
favoured by the collective expansion. Moreover, the nucleation can start now 
even above the critical line, when $P_{\rm H}<P_{\rm Q}$, and the standard 
theory would predict no growing bubbles. In principle,  
the phase separation can start as early as the metastable H state 
appears in the thermodynamic potential, and a stable interface between the 
phases may exist. 

Using Eqs. (\ref{busu}) and (\ref{kin}) one can determine  the critical 
bubble radius $R_c$, corresponding to the top of the potential 
barrier in $\Delta\Omega(R)$. The condition 
$\partial_R \Delta\Omega=0$ leads to a cubic equation for $R_c$.
When $H\to 0$ the kinetic term vanishes and this equation gives
a standard Laplace formula for the critical bubble \cite{Kap}.  
However for realistic parameters (see below) the kinetic term dominates.
In particular, in the vicinity of the critical line, when 
$P_{\rm H}\approx P_{\rm Q}$, one can consider  the
bulk term perturbatively. Then one obtains
\begin{equation} \label{crit}
R_c=\left(\frac{4\gamma}{\Delta {\cal E} H^2}\right)^{1/3}\left[
1-\frac{P_{\rm H}-P_{\rm Q}}{3\left(2\gamma^2\Delta {\cal E}
    H^2\right)^{1/3}}\right]~.
\end{equation}
The bubbles with $R>R_c$ will expand further while those with $R<R_c$ 
will eventually shrink. From the above consideration one should conclude that 
in a rapidly expanding system an appreciable amount of nucleation bubbles and
even empty cavities will be created already above the critical line. 

The bubble formation and growth will also continue below the critical line. 
Previously formed bubbles will now grow faster due to increasing pressure 
difference, $P_{\rm H}-P_{\rm Q}>0$, between the two phases. 
It is most likely  that the conversion of Q matter on the bubble boundary
is not fast enough to saturate the H phase. Therefore, a fast expansion  
may lead to a deeper cooling of the H phase inside the bubbles compared to 
the surrounding Q matter. Strictly speaking, such a system cannot be
characterized by the unique temperature. At some stage H bubbles 
will percolate, and the topology of the system will change. Now isolated 
regions of the Q phase (Q droplets) will be surrounded by the undersaturated 
vapor of the H phase. 
 
Standard thermodynamical concepts cannot be used in this non-equilibrium 
situation. However, the characteristic droplet size can be estimated by 
applying the energy balance consideration first
proposed by Grady \cite{Gra,Hol} in the study of dynamical fragmentation. 
The idea is that the fragmentation of expanding matter is a local 
process minimizing the sum of surface and kinetic (dilational) energies 
per fragment volume. The predictions of
this simple model are in reasonable agreement with molecular 
dynamics simulations \cite{Hol,Bli} and experimental data on dynamical 
fragmentation of fluids and solids (see e.g. \cite{Gra,Hol,Toe}). 
As shown in Ref. \cite{Mis}, this prescription works fairly well also 
for multifragmentation of expanding nuclei, where the 
standard statistical approach fails. 
 
Let us imagine an expanding spherical Q droplet embedded in the background 
of the dilute H phase. In the droplet rest frame the change of 
thermodynamic potential compared to the uniform H phase is given by the same 
expression (\ref{delta}) but with indexes H and Q interchanged. The kinetic
term is positive now. According to the Grady's prescription, the 
characteristic droplet radius, $R^\ast$, can be determined by minimizing 
\begin{equation}
\left(\frac{\Delta\Omega}{V}\right)_{droplet}=
-\left(P_{\rm Q}-P_{\rm H}\right)+\frac{3\gamma}{R}
+\frac{3}{10}\Delta {\cal E} H^2 R^2~. 
\end{equation}
It is worth noting that the collective kinetic energy term acts here as an 
effective long-range potential, similar to the Coulomb potential in nuclei.
Since the bulk term does not depend on $R$ the minimization condition 
constitutes the balance between the collective kinetic energy and interface 
energy. This leads to an optimum droplet radius
\begin{equation}  \label{R}
R^{\ast}=\left(\frac{5\gamma}{\Delta {\cal E} H^2}\right)^{1/3}.
\end{equation}
It should be noticed that this radius is expressed through the same 
combination of model parameters as the critical bubble radius at
$P_{\rm H}\approx P_{\rm Q}$, 
Eq. (\ref{crit}), but with a slightly bigger numerical coefficient.
This suggests that in vicinity of the critical line the H and Q phases occupy 
roughly equal fractions of the total volume. 
Fast expansion (large $H$) may lead to very small droplets. 
This mixed state of matter is far from thermodynamical equilibrium because
of the excessive interfacial energy and undersaturation of the H phase. 
One can say that the metastable Q matter is torn apart by a 
mechanical strain associated with the collective expansion. 
This has a direct analogy with the fragmentation of pressurized fluids 
leaving nozzles \cite{Bli,Toe}. In a similar way, splashed water forms
droplets which have nothing to do with the equilibrium liquid-gas phase 
transition.

At ultrarelativistic collision energies associated with RHIC and LHC 
experiments, the expansion will be very anisotropic, with its strongest
component along the beam direction. 
Applying the same consideration for the anisotropic flow, one can see that 
the characteristic size of inhomogeneities in each direction is determined 
by the respective Hubble constant acting in this direction. Thus, the 
resulting structures will have smaller size in the direction of stronger 
flow. Therefore, in the case of strong one-dimensional expansion the 
inhomogeneities associated with the phase separation will rearrange
themselves  into pancake-like slabs of Q matter layered by the dilute H 
phase. The characteristic width of the slab is given by Eq. (\ref{R}) with 
a slightly different geometrical factor. At a later stage the slabs will 
further fragment into smaller droplets. In general one should expect the 
emergence of complicated multi-fractal structures. 

The driving force for expansion is the pressure gradient, 
$\nabla P=c_s^2\nabla{\cal E}$, which depends on the sound velocity in the 
matter, $c_s$. In the vicinity of the critical line one may expect a ``soft 
point'' \cite{Shu,Ris} where $c_s$ is smallest and the ability of 
matter to generate the collective expansion is minimal (small $H$). 
If the initial state of the Q phase is close to this point, 
the primordial bubbles or droplets will be biggest. 
Increasing initial pressure will result in a faster expansion
and smaller droplets. For numerical estimates I choose two values of the 
Hubble constant: $H^{-1}$=20 fm/c to represent the slow expansion from the 
soft point \cite{Shu} and $H^{-1}$=6 fm/c for the fast expansion \cite{Zab}.

One should also specify two other parameters, $\gamma$ and 
$\Delta{\cal E}$. The surface tension $\gamma$ is a subject of debate at 
present. Lattice simulations indicate that at the critical point it could 
be as small as a few MeV/fm$^2$. However, for the  non-equilibrium situation
discussed here the values of 10-20 MeV/fm$^2$, which follow from 
effective chiral models, should be more appropriate. As a compromise, 
the value $\gamma=10$ MeV/fm$^2$ is used below. It is clear that 
$\Delta{\cal E}$
should be close to the latent heat of the transition, i.e. about 
$0.5\div 1$ GeV/fm$^{3}$. One can also estimate $\Delta{\cal E}$ by  
realizing that nucleons and heavy mesons are the smallest 
droplets of the Q phase.  For estimates I  take $\Delta{\cal E}=0.5$ 
GeV/fm$^3$, i.e. the energy density inside the nucleon. 
Substituting these values in Eq. (\ref{R}) one gets $R^\ast$=3.4 fm for 
$H^{-1}$=20 fm/c and $R^\ast$=1.5 fm for 
$H^{-1}$=6 fm/c. 

In the lowest-order approximation the characteristic droplet mass is 
$M^\ast\approx\Delta{\cal E}V$.
For spherical and slab-like droplets one gets respectively
\begin{equation} \label{mass}
M^{\ast}_{sp} \approx \frac{20\pi}{3}\frac{\gamma}{H^2}~,~~~
M^{\ast}_{sl} \approx 2S\left(\Delta{\cal E}\right)^{2/3}
\left(\frac{3\gamma}{H^2}\right)^{1/3}~,
\end{equation} 
where $S$ is the slab transverse area. It is interesting to note that in
this approximation $M^\ast_{sp}$ is independent of $\Delta{\cal E}$. 
For the two values of $R^*$ given above $M^\ast_{sp}$ is $\sim$100 GeV and 
$\sim$10 GeV, respectively. The slab-like structures would have even larger 
mass, since $S$ could be of order of the transverse system size. 
Using the minimum information 
principle one can show \cite{Hol,Mis} that the distribution of droplets 
should follow an exponential law, $\exp{\left(-{M \over M^{\ast}}\right)}$. 
Therefore, with 1$\%$ probability one
can  find droplets as heavy as $5M^{\ast}$. 

After separation, the droplets recede from each other according to the 
global Hubble expansion, predominantly along the beam direction. Therefore,
their center-of-mass 
rapidities are in one-to-one correspondence with their spatial positions.
Presumably they will be distributed more or less evenly between 
the target and projectile rapidities. 
At this late stage it is unlikely that 
the thermodynamical equilibrium will be re-established 
between the Q and H phases or within the H phase alone. If this were to 
happen, the final H phase would be uniform, and thus
there would be no traces of the mixed phase in the final state. 

The final fate of individual droplets depends on their sizes and on 
details of the equation of state. Due to the counter-acting pressure of the H
phase and additional Laplace pressure 
the initial expansion will slow down. In smaller droplets the expansion and 
cooling may even reverse to the contraction and reheating. The conversion of 
Q droplets into the H phase may proceed through formation of a deflagration 
front \cite{Ris} or evaporation of hadrons from the surface \cite{Alf}.
Bigger droplets may expand further until they enter the region of
spinodal instability. 
As shown in Ref. \cite{Cse}, the characteristic time of the 
``rolling down'' process is rather short, $\sim 1$ fm/c, so that the 
Q droplets will be converted rapidly into the H phase. 
The energy released in this process can be transferred partly into the 
collective oscillations of the ($\sigma,{\bf \pi}$) fields. 
Numerical simulations \cite{Aba,Sca} show that these oscillations persist 
for a long time and give rise to soft pion radiation. One should also expect 
the formation of Disoriented Chiral Condensates (DCC) in the voids between 
droplets. 


Since rescatterings in the dilute H phase are rare, most hadrons produced 
from individual droplets will go directly into detectors.  One may guess that
the number of produced hadrons is proportional to the droplet mass. Each 
droplet will give a bump in the hadron rapidity distribution around its 
center-of-mass 
rapidity \cite{Cse}. If emitted particles have a Boltzmann spectrum, the 
width of the bump will be $\delta y \sim 2\sqrt{T/m}$, where $T$ is the 
droplet temperature and $m$ is the particle mass. At $T\sim 100$ MeV this 
gives $\delta y\approx 2$ for pions and $\delta y\approx 1$ for nucleons. 
These spectra might be slightly modified by the residual expansion of 
droplets and their transverse motion. The resulting rapidity distribution 
in a single event will be a superposition of 
contributions from different droplets, and therefore it will exhibit strong 
non-statistical fluctuations. The fluctuations will be more pronounced 
if primordial droplets are big, as expected in the vicinity of the soft point.
If droplets as heavy as 100 GeV are formed, each of them will produce up to 
$\sim$300 pions within a narrow rapidity interval, $\delta y\sim 1$. 
Such bumps can be easily resolved and analyzed. Critical
fluctuations of similar nature were discussed recently in Ref. \cite{Ant}.
  
Some unusual events produced by high-energy cosmic nuclei have been already 
seen by the JACEE collaboration \cite{JACEE}. Unfortunately, they are very few 
and it is difficult to draw definite conclusions by analyzing them. 
We should be prepared to see thousands of such events in the future RHIC and 
LHC experiments.
It is clear that the nontrivial structure of the hadronic spectra will be 
washed out to a great extent when averaging over many events. Therefore, more 
sophisticated methods of the event sample analysis should be used. 
The simplest one is to  search for non-statistical fluctuations in the 
hadron multiplicity distributions measured in a fixed rapidity bin
\cite{Tan}. One can also study the correlation of multiplicities in 
neighbouring rapidity bins, bump-bump correlations etc. 
Such standard methods as intermittency and commulant moments \cite{Ant}, 
wavelet transforms \cite{Suz}, HBT interferometry \cite{Hei} can also be
useful. All these studies should be done at 
different collision energies to identify the phase transition threshold.
The predicted dependence on the Hubble constant and the geometry of reaction,
Eq. (\ref{mass}), can be checked in collisions with different ion masses and 
impact parameters.

One should bear in mind two important points. First, if the expansion 
trajectory goes close to the (tri)critical point, both $\gamma$ and
$\delta{\cal E}$ will tend to zero and the critical fluctuations will be 
less pronounced. Second, if a first order phase transition is 
possible only in the baryon-rich matter, then the Q droplets should have 
much higher baryon density than the hadronic phase \cite{Jac}. In this case 
one should expect strong non-statistical fluctuations in the distribution 
of the net baryon charge. 


In conclusion, simple arguments based on the homogeneous nucleation 
picture and the energy balance consideration demonstrate that 
a first order phase transition in rapidly expanding matter should proceed 
through a non-equilibrium stage when the metastable phase fragments into
droplets. If QCD matter undergoes a first order phase transition, it will 
manifest itself in relativistic heavy-ion collisions by the formation of 
droplets of quark-gluon plasma. The primordial droplets should be 
biggest in the vicinity of the soft point where the expansion is slowest.
The fragmentation of plasma might be accompanied by the formation of 
multiple DCC domains and enhanced soft-pion radiation. 
Subsequent hadronization of QGP droplets will lead to large non-statistical 
fluctuations in the hadron rapidity spectra in individual events. 
All these novel phenomena can only be detected through
dedicated event-by-event analysis of experimental data.

The author is grateful to J.P. Bondorf and A.D. Jackson for many fruitful 
discussions. I thank Agnes Mocsy for preparing the figure. Discussions 
with G. Carter, D. Diakonov, J.J. Gaardh{o}je, L. McLerran, R. Mattiello, 
L.M. Satarov and E.V. Shuryak are greatly appreciated. 
I thank the Niels Bohr Institute for kind hospitality and financial support.

\newpage

\begin{figure}
\vspace{-2cm}
\epsfxsize=16cm
\epsffile{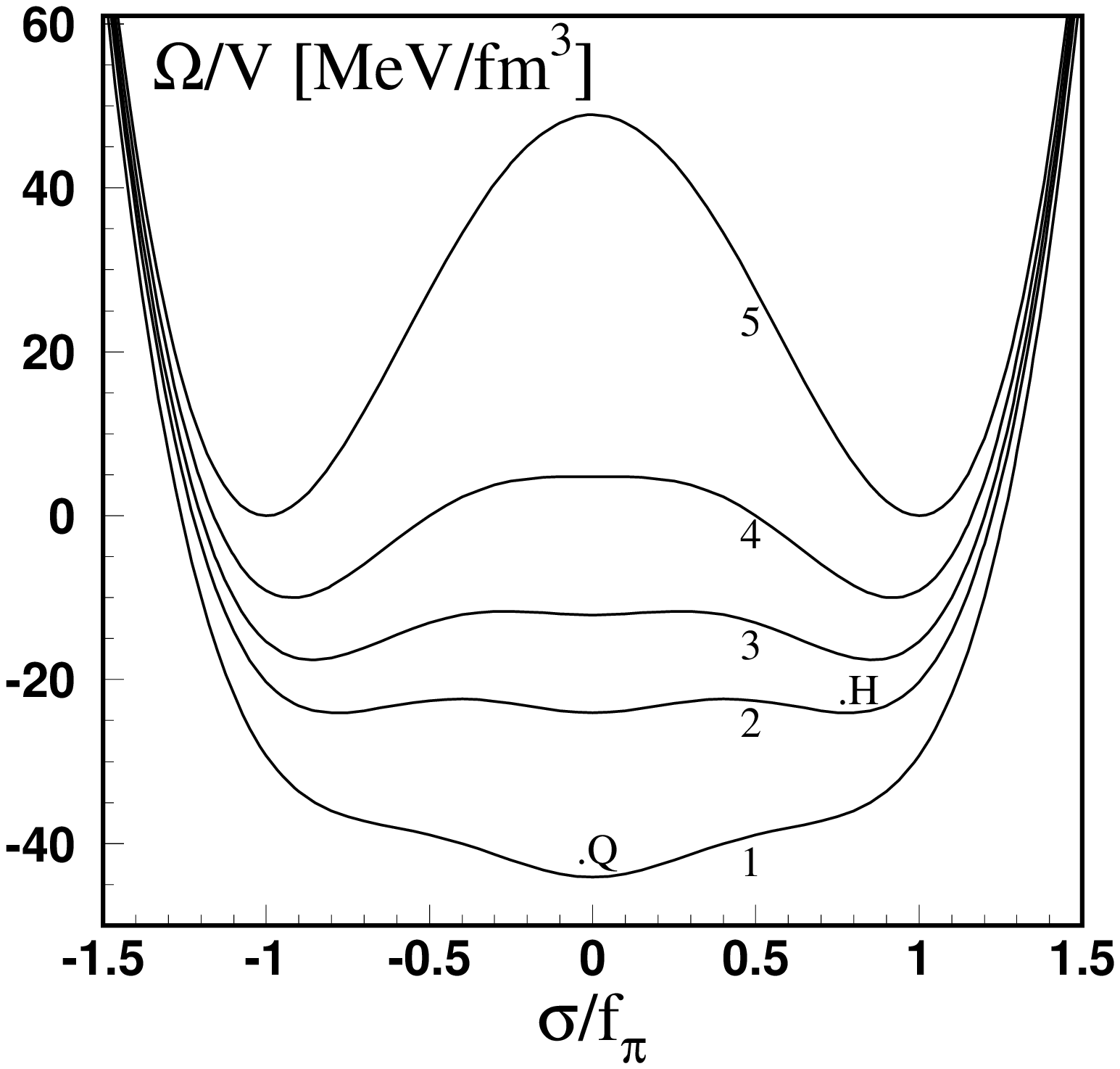}
\vspace{-3cm}
\caption{Schematic view of the effective thermodynamic potential per volume
$\Omega/V$ as a function of the order parameter field $\sigma$ at
${\bf \pi}=0$, as predicted by the linear $\sigma$-model in the chiral limit 
$m_\pi=0$ [1]. The curves from bottom to top correspond to the 
different stages of the isentropic expansion of homogeneous matter starting 
from $T$=100 MeV and $\mu$=750 MeV (curve 1). The upper 
curve 5 is the vacuum potential. The other curves are discussed in the text.}
\end{figure}
 

\begin{thebibliography}{99}

\bibitem{Moc} L.P. Csernai, I.N. Mishustin and A. Mocsy, Heavy Ion Physics,
{\bf 3}, 151 (1996);\\ A. Mocsy, M.Sc. thesis, University of Bergen, 1996. 

\bibitem{Raj} J. Berges and K. Rajagopal, Nucl. Phys. {\bf B} (in press);
hep-ph/9804233.

\bibitem{Jac} M.A. Halasz, A.D. Jackson, R.E. Shrock, M.A. Stephanov and 
J.J.M. Verbarshot, Phys. Rev. {\bf D58}, 096007 (1998).

\bibitem{Car} G. Carter and D. Diakonov, Nucl. Phys. {\bf A} (in press); 
hep-ph/9807219. 

\bibitem{Ste} M. Stephanov, K. Rajagopal and E. Shuryak, Phys. Rev. Lett. 
{\bf 81}, 4816 (1998).

\bibitem{Sat} I.N. Mishustin, L.M. Satarov, H. Stoecker and W. Greiner,
hep-ph/9812319.

\bibitem{Bra} P. Braun-M\"unzinger and J. Stachel, 
Nucl. Phys. {\bf A638}, 3c (1998).

\bibitem{Kap} L.P. Csernai, J.I. Kapusta, Phys. Rev. Lett. {\bf 69}, 737 
(1992); Phys. Rev. {\bf D46}, 1379 (1992).

\bibitem{Zab} E.E. Zabrodin, L.V. Bravina, L.P. Csernai, H. Stoecker and 
W. Greiner, Phys. Lett. {\bf B423}, 373 (1998). 

\bibitem{Gra} D.E. Grady, J. Appl. Phys. {\bf 53}(1), 322 (1981).

\bibitem{Hol} B.L. Holian and D.E. Grady, Phys. Rev. Lett. {\bf 60}, 1355 
(1988).

\bibitem{Bli} J.A. Blink and W.G. Hoover, Phys. Rev. {\bf A32}, 1027 (1985).

\bibitem{Toe} H.Buchenau et al., 
J. Chem. Phys. {\bf 92}, 6875 (1990). 


\bibitem{Mis} I.N. Mishustin, 
Nucl. Phys. {\bf A630}, 111c (1998).

\bibitem{Shu} C.M. Huang, E.V. Shuryak, Phys. Rev. Lett. {\bf 75}, 4003 
(1995); Phys. Rev. {\bf C57}, 1891 (1998).  

\bibitem{Ris} D. Rischke, M. Gyulassy, Nucl. Phys. {\bf A597}, 701 (1996); 
{\bf A608}, 479 (1996).


\bibitem{Alf} M. Alford, K. Rajagopal and F. Wilczek, Phys. Lett. {\bf B422},
247 (1998).

\bibitem{Cse} L.P. Csernai, I.N. Mishustin, Phys. Rev. Lett. {\bf 74}, 5005 
(1995).

\bibitem{Aba} A. Abada, M. Birse, Phys. Rev. {\bf D55}, 6885 (1987).

\bibitem{Sca} I.N. Mishustin, O. Scavenius, 
Nucl. Phys. {\bf A638}, 519c (1998).


\bibitem{Ant} N.G. Antoniou, Nucl. Phys. {\bf B71}, 307 (1999).

\bibitem{JACEE} T.H. Barnett et al., Phys. Rev. Lett. {\bf 50}, 2062 (1983).

\bibitem{Tan} M.J. Tannenbaum and E802 Collaboration, Phys. Rev. {\bf C52}, 
2663 (1995)

\bibitem{Suz} N. Suzuki, M. Biyajima and A. Ohsawa, hep-ph/9503403. 

\bibitem{Hei} H. Heiselberg, A.D. Jackson, hep-ph/9809013.

.


\end{thebibliography}
\end{document}